\documentclass[pdflatex,sn-mathphys-num]{sn-jnl}

\usepackage{graphicx}%
\usepackage{multirow}%
\usepackage{amsmath,amssymb,amsfonts}%
\usepackage{amsthm}%
\usepackage{mathrsfs}%
\usepackage[title]{appendix}%
\usepackage{xcolor}%
\usepackage{textcomp}%
\usepackage{manyfoot}%
\usepackage{booktabs}%
\usepackage{algorithmicx}%
\usepackage{algpseudocode}%
\usepackage{listings}%

\usepackage[justification=centering]{caption}

\usepackage[linesnumbered,ruled,vlined]{algorithm2e}
\SetKwProg{Fn}{Function}{:}{}
\SetEndCharOfAlgoLine{}

\theoremstyle{thmstyleone}%
\theoremstyle{thmstyletwo}%

\theoremstyle{thmstylethree}%
\newtheorem{definition}{Definition}%

\begin{document}

\title[Article Title]{QDBFT: A Dynamic Consensus Algorithm for Quantum-Secured Blockchain}


\author[2]{\fnm{Fei} \sur{Xu}}
\equalcont{These authors contributed equally to this work.}

\author[1,2]{\fnm{Cheng} \sur{Ye}}
\equalcont{These authors contributed equally to this work.}

\author[2]{\fnm{Jie} \sur{OuYang}}

\author[2]{\fnm{Ziqiang} \sur{Wu}}

\author[1,2]{\fnm{Haoze} \sur{Chen}}

\author[1,2]{\fnm{An} \sur{Hua}}

\author[2]{\fnm{Meifeng} \sur{Gao}}

\author[2]{\fnm{Qiandong} \sur{Zhang}}

\author[1,2]{\fnm{Minghan} \sur{Li}}

\author[1,2]{\fnm{Feilong} \sur{Li}}

\author*[1,2]{\fnm{Yajun} \sur{Miao}}\email{miaoyajun@qtict.com}

\author*[1,2]{\fnm{Wei} \sur{Qi}}\email{qiwei@qtict.com}

\affil[1]{\orgname{CAS Quantum Network Co., Ltd.}, \orgaddress{\city{Shanghai}, \country{China}}}

\affil[2]{\orgname{Anhui CAS Quantum Network Co., Ltd.}, \orgaddress{\city{Hefei}, \country{China}}}


\abstract{The security foundation of blockchain system relies primarily on classical cryptographic methods and consensus algorithms. However, the advent of quantum computing poses a significant threat to conventional public-key cryptosystems based on computational hardness assumptions.
In particular, Shor’s algorithm can efficiently solve discrete logarithm and integer factorization problems in polynomial time, thereby undermining the immutability and security guarantees of existing systems. Moreover, current Practical Byzantine Fault Tolerance (PBFT) protocols, widely adopted in consortium blockchains, suffer from high communication overhead and limited efficiency when coping with dynamic node reconfigurations, while offering no intrinsic protection against quantum adversaries.

To address these challenges, we propose QDBFT, a quantum-secured dynamic consensus algorithm, with two main contributions: first,we design a primary node automatic rotation mechanism based on a consistent hash ring to enable consensus under dynamic membership changes, ensuring equitable authority distribution;  second, we integrate Quantum Key Distribution (QKD) networks to provide message authentication for inter-node communication, thereby achieving information-theoretic security in the consensus process. 
Experimental evaluations demonstrate that QDBFT achieves performance comparable to traditional PBFT while delivering strong resilience against quantum attacks, making it a promising solution for future quantum-secure decentralized infrastructures.}




\maketitle

\section{Introduction}\label{sec1}

As the fundamental infrastructure of distributed ledgers, blockchain technology ensures data immutability and traceability through cryptographic mechanisms and consensus protocols. In particular, the chained data structure anchored by cryptographic hash links, together with public auditability, provides strong resistance against tampering with historical records, while consensus protocols ensure the stable operation of the blockchain even in the presence of faulty nodes.

The security of traditional blockchain systems is based on the computational hardness assumptions underlying classical public-key cryptography(e.g., DSA \cite{1250416}, ECDSA \cite{johnson2001elliptic}, RSA \cite{rivest1978method}) and hash functions (e.g., SHA-256 \cite{pub2012secure}). However, the emergence of quantum computing  threatens this foundation. Shor’s algorithm \cite{shor1994algorithms}, with its ability to solve discrete logarithm and large integer factorization problems in polynomial time, enables the recovery of private keys in many existing signature schemes. Recent studies suggest that a quantum computer equipped with fewer than one million noisy qubits could break the widely deployed 2048‑bit RSA cryptosystem within approximately seven days \cite{gidney2025factor}. Meanwhile, Grover’s algorithm \cite{grover1996fast}, a well‑established quantum search technique, reduces the complexity of finding hash collisions from $O(2^n)$ to $O(2^{n/2})$, thereby undermining the collision resistance of hash functions and, consequently, threatening the immutability of blockchains. According to NIST projections, with the rapid progress of quantum technology, practical quantum computers are anticipated to become feasible within the coming decades \cite{moody2024pqs}.

Currently, consensus algorithms are primarily based on the Byzantine Fault Tolerance (BFT) mechanism, whose theoretical foundation stems from the Byzantine Generals' Problem proposed by Lamport et al. \cite{lamport2019byzantine}. The core challenge lies in achieving global consistency while tolerating malicious nodes that arbitrarily deviate from protocol rules. Although these algorithms can tolerate up to $f$ faulty nodes, their memberships are usually static, which imposes strict constraints on practical network environments. In contrast, dynamic consensus schemes attempt to relax this limitation by allowing flexible node participation, yet they often incur substantial communication overhead during membership updates. This observation underscores the importance of efficient node management, as schemes capable of accommodating dynamic membership with lower communication overhead are better positioned to achieve practical scalability.

Motivated by these issues, we investigate the quantum-resistant properties of existing blockchains and the optimization of communication overhead during dynamic node transitions in consensus algorithms.

\section{Related Work}\label{sec2}

Consensus mechanisms derived from the Byzantine Generals’ Problem are typically classified into three network communication models: synchronous, partially synchronous, and asynchronous. The resilience of each model to network latency directly determines the protocol’s security bounds and performance metrics. Early research focused on the synchronous model \cite{dolev1985bounds, dolev1982polynomial, katz2009expected, pease1980reaching}; however, the assumption of a strict upper bound on message delay limits its applicability in highly dynamic asynchronous environments. Concurrently, the FLP impossibility theorem \cite{fischer1985impossibility} demonstrates that the development of asynchronous network models is fundamentally constrained by the trade-off between security and liveness, which often requires the use of randomization mechanisms \cite{ben1983another, cachin2000random, liu2023flexible}.

The partially synchronous model \cite{dwork1988consensus} reconciles security and liveness by introducing the notion of Global Stabilization Time (GST). This model has become the most widely adopted communication framework for consortium blockchain consensus protocols \cite{castro1999practical, buchman2016tendermint, yin2019hotstuff}. Among its representative protocols, PBFT \cite{castro1999practical, castro1999authenticated} represented the first practical breakthrough in Byzantine Fault Tolerance under a partially synchronous setting. By employing a three-phase protocol (Pre-prepare, Prepare, and Commit) and supporting a fault tolerance scale of $N \geq 3f + 1$, PBFT provided the first viable BFT solution for real-world distributed systems. RBFT (Redundant BFT) \cite{buchman2016tendermint} enhances robustness and resistance against attacks by employing redundant phases to monitor primary node efficiency and utilizing a view-change mechanism; although it improves communication efficiency, primary node transitions still incur significant communication overhead. HotStuff \cite{yin2019hotstuff} introduces threshold signatures to transform the communication structure, where the primary node handles, aggregates, and forwards messages. However, dynamic events such as node joining, exiting, or going offline continue to pose communication challenges across these consensus schemes.

Regarding dynamic node membership, Duan et al. \cite{9833787} provided the first formal treatment, introducing a flexible formal framework and a set of security definitions while designing a dynamic BFT protocol. Tang et al. \cite{TANG2024107922} proposed a dynamically scalable BFT protocol for consortium blockchains that combines Distributed Key Generation (DKG) with BLS aggregate signatures to achieve dynamic node management. Li et al. \cite{LI2025103386} introduced DynaNet, a membership management framework built on BFT consensus, which supports arbitrary membership changes within a single consensus round and can be integrated into most synchronous BFT protocols to provide membership-change capabilities. Nevertheless, these dynamic BFT schemes still incur significant communication overhead during membership transitions and do not address the challenges of quantum resistance.

To remain secure against quantum threats, several blockchain architectures have adopted Post-Quantum Cryptography (PQC) algorithms \cite{QRL, Abelian, TheTangle, gao2018secure}. However, compared with conventional public-key cryptosystems, PQC introduces performance bottlenecks such as larger key sizes and reduced computational efficiency, posing challenges to simultaneously achieving quantum-resistant security and high operational efficiency.

Several studies \cite{kiktenko2018quantum, sun2019towards} have leveraged Quantum Key Distribution (QKD) to enhance key security and build information-theoretically secure blockchain architectures. Current research on QKD-based blockchains primarily focuses on a hybrid architecture consisting of a quantum security layer and a classical consensus layer. In such designs, the QKD network serves as the quantum physical security layer, providing information-theoretic security (ITS) for inter-node communication through techniques such as message authentication. Kiktenko et al. \cite{kiktenko2018quantum} proposed a quantum-secured blockchain architecture in 2017 comprising a QKD layer and a classical communication layer. Their scheme employs Toeplitz hashing \cite{toeplitz} for message authentication to replace digital signatures in public-key infrastructures, ensuring consistency via continuous inter-node message exchange. However, due to its high communication complexity, this scheme is impractical in large-scale networks. Sun et al. \cite{sun2019towards} combined the information-theoretically secure signature scheme by Amiri et al. \cite{amiri2018efficient} with Toeplitz hashing to verify message sources within the consensus mechanism. Nevertheless, their approach does not account for exception handling when the primary node is malicious, resulting in low efficiency under primary-node-fault scenarios. Singh et al. \cite{10486256} suggested using QKD to enhance blockchain security against quantum threats and validated their approach using a QKD network simulator. Faruk et al. \cite{10.1007/978-3-031-56728-5_22} proposed a two-stage scheme integrating Hyperledger Fabric with QKD protocols, employing a permissioned blockchain for quantum device identity authentication and securing the parameter reconciliation process to strengthen QKD systems against Man-in-the-Middle (MitM) attacks.

Currently, research on QKD-based consensus in quantum-secured blockchains remains largely confined to modifying digital signatures within static consensus schemes. There is still a lack of comprehensive exploration into the integration of quantum-resistant security within dynamic BFT frameworks. Moreover, existing QKD-based approaches suffer from low efficiency, limiting their practicality in complex, large-scale scenarios.

To address the challenges of high communication overhead during membership transitions, low signature verification efficiency, and the lack of quantum resistance in existing dynamic BFT schemes, we propose QDBFT (Quantum Dynamic-node Byzantine Fault Tolerance), a flexible and quantum-secured dynamic consensus mechanism. By leveraging an automatic node rotation mechanism together with Quantum Key Distribution (QKD) networks, QDBFT enables low-overhead random node transitions during block generation and implements quantum-secure message authentication. Our main contributions can be summarized as follows:

First, an automatic node rotation mechanism based on a consistent hash ring is designed to achieve consensus. By maintaining and utilizing the latest configuration table $T_v$, the system ensures that all membership changes—including node join requests, voluntary exits, and unresponsive events—are incorporated into the consensus process and agreed upon collectively. This mechanism also addresses the issue of centralized block-proposing authority by enabling an equitable distribution of transaction validation rights.

Second, we introduce a quantum-secure authentication signature algorithm built upon the information-theoretically secure keys provided by QKD networks. This equips nodes with message authentication capabilities that remain resilient against quantum computing threats.

\section{Preliminaries}\label{sec3}

\subsection{Quantum Key Distribution (QKD)}\label{subsec31}
Quantum Key Distribution (QKD) is a cryptographic protocol based on the principles of quantum mechanics \cite{heisenberg_uber_1927, wootters_single_1982}. By transmitting quantum states over a quantum channel, QKD enables the detection of any interference introduced by an eavesdropper. This allows communicating parties to establish a shared key with information-theoretic security (ITS), as exemplified by the BB84 protocol \cite{BENNETT20147}.

A QKD network is a secure infrastructure for key distribution built upon QKD technology. Its primary purpose is to extend point-to-point QKD links into a scalable system that supports on-demand distribution, relaying, and management of secure keys across distant nodes. In the quantum-secured blockchain considered in this paper, the QKD network serves as a foundational security layer, establishing shared pools of random symmetric keys between every pair of nodes in the consensus cluster. These keys enable quantum-secure authentication, thereby forming the basis for quantum-resistant guarantees in the proposed QDBFT consensus protocol.

The workflow of a QKD network can be summarized in three stages:
\begin{itemize} 
\item \textbf{Key Distribution:} 
Adjacent nodes establish physical links via QKD devices and execute protocols such as BB84 to negotiate symmetric keys. For non-adjacent nodes, end-to-end shared keys are established through trusted-node relaying.
\item \textbf{Key Storage and Management:} 
Generated keys are securely stored in each node’s local key pool, which supports comprehensive lifecycle management, including identification, renewal, and updates.
\item \textbf{Key Usage:} 
During consensus communication, nodes retrieve pre-shared keys from the key pool to generate quantum-secure authentication signatures, thereby ensuring both message integrity and authenticity.
\end{itemize}

Compared with traditional public-key cryptosystems that rely on computational hardness assumptions, QKD offers the distinct advantage of providing information-theoretic security. Even adversaries equipped with immense computational resources—including future quantum computers—cannot compromise the keys generated by QKD.

\subsection{Practical Byzantine Fault Tolerance (PBFT)}

Practical Byzantine Fault Tolerance (PBFT) \cite{castro1999practical} is the first State Machine Replication (SMR) protocol to demonstrate practicality in a partially synchronous network model. Its primary goal is to achieve consistent system state through collaboration among nodes in a distributed environment, tolerating up to $f$ malicious (Byzantine) nodes as long as the total number of nodes satisfies $N \geq 3f+1$.

During normal operation (i.e., when no view change occurs), the PBFT protocol proceeds through three main phases: Pre-prepare, Prepare, and Commit.
\begin{itemize}
\item \textbf{Request:} The client sends a signed request message to the primary node, requesting the execution of a specific operation.
\item \textbf{Pre-prepare:} 
The primary node assigns a sequence number to the valid client request and broadcasts a Pre-prepare message to all replicas. This message includes the sequence number, the current view number, and the digest of the request.
\item \textbf{Prepare:}
After validating the Pre-prepare message, each replica broadcasts a Prepare message to all other replicas. A replica enters the "Prepared" state once it has received $2f$ matching Prepare messages from distinct nodes (for a total of $2f+1$ including itself).

\item \textbf{Commit:} Replicas in the "Prepared" state broadcast a Commit message. A replica applies the request to its state machine and executes the operation once it has collected $2f+1$ valid Commit messages, indicating that a majority of nodes agree on the request sequence.

\item \textbf{Reply:} After executing the operation, replicas return the result to the client. The client confirms the successful execution upon receiving $f+1$ identical replies.
\end{itemize}

Through this three-phase protocol, PBFT ensures that all honest nodes execute requests in the same order, thereby maintaining state machine consistency. If the primary node behaves abnormally or maliciously, a view change process is triggered to elect a new primary.

\section{QDBFT}\label{sec4}
\subsection{Notations}
The definitions of the primary notations used in the description of the QDBFT algorithm are summarized in Table \ref{tab1}.

\begin{table}[htbp]
\centering
\caption{Notations and Descriptions}
\label{tab1}
\begin{tabular}{ll}
\toprule
\textbf{Notation} & \textbf{Description} \\
\midrule
$H^\prime(k,m)$ & Hash-based message authentication function \\
$m$ & Message to be signed \\
$M$ & Set of messages to be signed in the system \\
$N$ & Set of consensus nodes \\
$T_v$ & Node configuration table for version $v$ \\
$T^u_v$ & Updated configuration table \\
$H(pb)$ & Hash value of the parent block \\
$H_{32}(\cdot)$ & Hash function for the consistent hash ring mechanism \\
$i, j, l \in N$ & Individual consensus nodes \\
$Z$ & Number of virtual nodes on the hash ring \\
$t, T$ & Timestamps \\
$D(\cdot)$ & Message digest algorithm \\
$\Delta T$ & Designated time interval \\
$c_{pk}$ & Public key of the client \\
$nodes_{pk}$ & Set of public keys of consensus nodes \\
$\alpha$ & Authentication tag of a message \\
$G$ & Number of proposals in a block \\
\bottomrule
\end{tabular} \vspace{2mm} 
\end{table}

The function $H^\prime(k,m)$ represents the hash-based message authentication, implemented via HMAC or the Toeplitz algorithm. The set $N$ represents the consensus nodes, whose composition evolves dynamically through join requests, voluntary departures, or unresponsive events. Any such membership change necessitates an update to the configuration table $T_v$, and the most recent consensus-validated version $T^u_v$ must be incorporated into the request messages across all stages.The parent block hash $H(pb)$ has a length determined by the security requirements of the protocol, and $H_{32}(\cdot)$ is used to map this hash value onto the \textbf{consistent hash ring}. Timestamps $t$ and $T$ represent the message transmission time and the block commitment time, respectively. To accommodate large message sizes, the digest algorithm $D(\cdot)$ is employed for efficient communication. In addition, $\Delta T$ defines the threshold for node timeout detection, while $\alpha$ provides identity authentication and ensures message integrity during node–client interactions. Finally, $G$ specifies the number of proposals per block, directly affecting both storage overhead and client latency.

\subsection{Carousel: The Automatic Node Rotation Mechanism}
Karger et al. \cite{10.1145/258533.258660} introduced the consistent hash ring to address server load balancing challenges.  Building upon the concept of a virtual hash ring, we propose Carousel, an automatic node rotation mechanism designed to ensure the equitable distribution of authority among nodes, as illustrated in Fig. 1 and Definition 1.The primary node selected through the Carousel mechanism is granted proposal right for the new block. To guarantee that each node has an equal probability of being chosen when the node count is small, multiple virtual nodes are uniformly assigned to each physical node. The mechanism automatically handles membership dynamics—including node joins, voluntary departures, and unresponsive events—through version updates of the configuration table $T_v$. The updated configuration table, denoted as $T^u_v$, takes effect only after consensus validation, namely upon the reception of $2f$ consistent messages.

\begin{figure}
    \centering
    \includegraphics[scale=0.5]{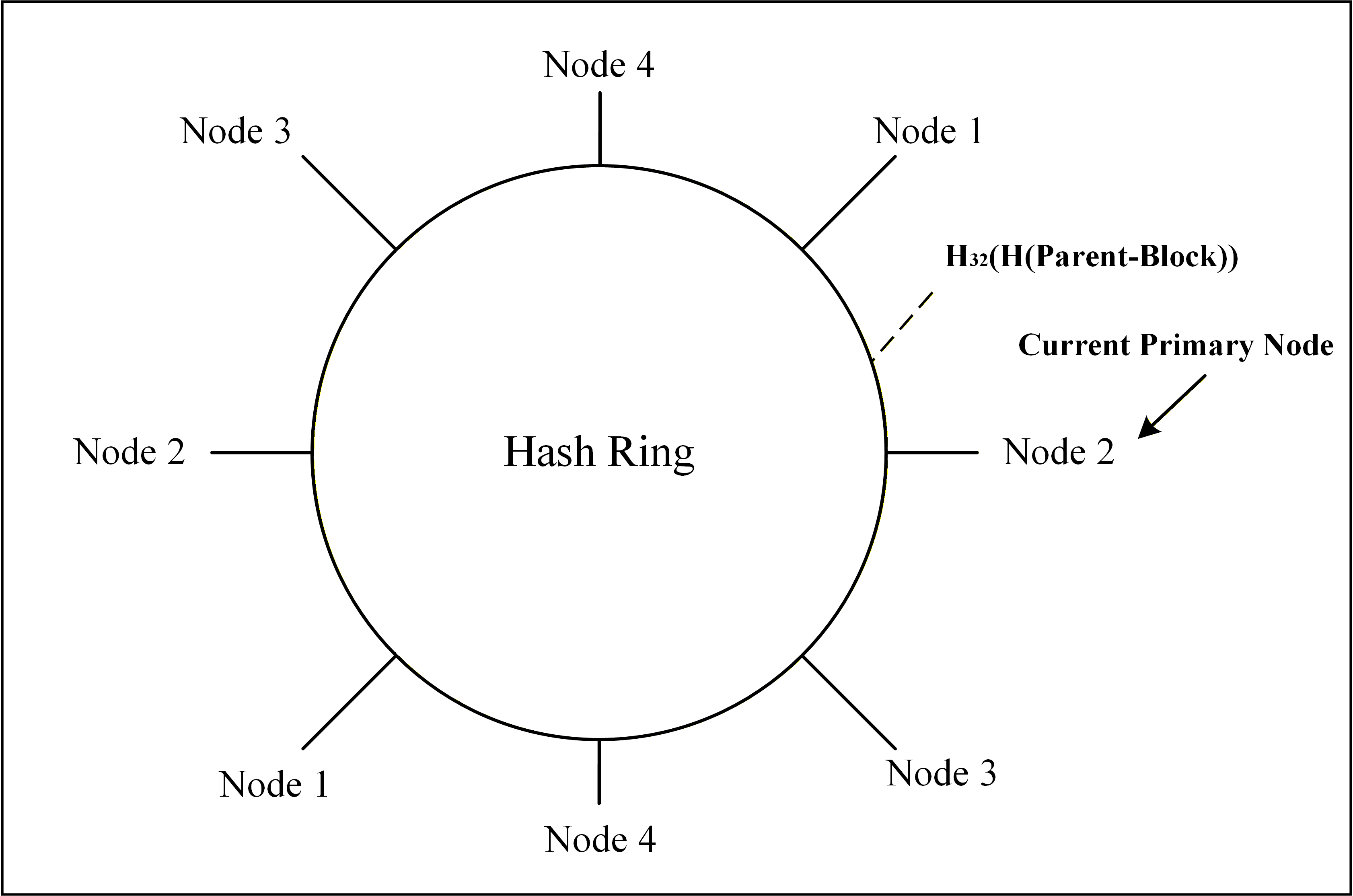}
    \caption{The structure of the Hash Ring and node mapping}
    \label{fig:hash_ring}
\end{figure}

\begin{definition}
Carousel: The Automatic Node Rotation Mechanism.

\noindent The system comprises a set of $N$ nodes with ordered identifiers. A 32-bit consistent hash ring (with a value range of $[0, 2^{32}-1]$) is utilized to assign equidistant hash points to each node, generating a configuration table $T_v: \{id_i, H_z\}$ (where $0 < i \leq N, 0 < z \leq Z$), and $v$ denotes the configuration version. This table is collectively maintained by the node set. The hash value of the parent block is mapped onto the 32-bit hash ring; the node corresponding to the first hash point encountered in a clockwise direction from this mapping is designated as the current primary node. Upon the successful consensus and commitment of a new block, the proposal and commitment rights are transferred to the next primary node determined by the same hash ring calculation.
\end{definition}

\subsection{QKD-based Message Authentication} 
QKD networks address security challenges in key distribution by leveraging quantum technology to establish an information-theoretically secure symmetric key $k$ between two endpoints. These symmetric keys serve as the foundation of the authentication algorithm within the QDBFT framework. In contrast to classical digital signatures, which rely on the computational hardness of mathematical problems, the proposed QKD-based message authentication algorithm exploits the intrinsic randomness and physical security of QKD-generated keys to ensure both message integrity and source authentication.

\begin{algorithm}
\caption{QKD-based Message Authentication Algorithm}
\label{alg:quantum_authentication}
\SetKwFunction{FSign}{message\_authentication\_code\_broadcast}
\SetKwFunction{FVerify}{message\_authentication\_code\_verification}

\nonfrenchspacing
\SetKwProg{Fn}{Function}{:}{}

\Fn{\FSign{$m, \{k_{l,e} \mid 1 \leq e \leq N, e \neq l\}$}}{
    \KwIn{$m$: message to be authenticated; $\{k_{l,e}\}$: pre-shared symmetric keys between node $l$ and all other nodes $e$.}
    \KwOut{$m_\alpha$: the authentication bundle.}
    $D(m) \leftarrow \text{compute digest of } m$\;
    \For{$e \leftarrow 1$ \KwTo $N, e \neq l$}{
        Compute $H'(k_{l,e}, D(m))$\;
    }
    Construct $m_\alpha \leftarrow \{D(m), \{H'(k_{l,e}, D(m)) \mid 1 \leq e \leq N, e \neq l\}\}$\;
    \textbf{Broadcast} $m_\alpha$\;
}

\vspace{0.5em}

\Fn{\FVerify{$m_\alpha, k$}}{
    \KwIn{$m_\alpha$: received authenticated message; $k$: the local symmetric key shared with the sender.}
    \KwOut{$1$ (Success) or $0$ (Failure).}
    Parse $m_\alpha$ to obtain $D(m)$ and the set of authentication tags $\{H'(k_{l,e}, D(m))\}$\;
    \For{$e \leftarrow 1$ \KwTo $N, e \neq l$}{
        \If{$H'(k, D(m)) == H'(k_{l,e}, D(m))$}{
            \Return $1$\;
        }
        \Else{
            \Return $0$\;
        }
    }
}
\end{algorithm}

\textbf{Key Distribution Phase:}  
Nodes within the set $N$ exchange symmetric keys via the QKD network and store them securely in local key pools. These symmetric keys are directional:
\begin{itemize}
\item $k_{i,j}$: utilized by node $i$ to provide authentication when sending data to node $j$.
\item $k_{j,i}$: utilized by node $j$ to provide authentication when sending data to node $i$.
\end{itemize}

\textbf{Authentication Tag Generation Phase:}  
As detailed in the \textit{message\_authentication\_code\_broadcast} function of Algorithm~\ref{alg:quantum_authentication}, 
the source node $l$ processes the message $m$ to obtain the message digest $D(m)$. Using a one-time hash algorithm $H^\prime$ and the pre-shared keys $k_{l,e}$ ($1 \leq e \leq N, \; e \neq l$), 
node $l$ generates the authentication bundle 
$m_\alpha := \{D(m), H^\prime(k_{l,e}, D(m)) \mid 1 \leq e \leq N, \; e \neq l\}.$

Subsequently, node $l$ broadcasts both the message $m$ and its authentication bundle $m_{\alpha}$ to all verification nodes $e$ via authenticated channels.  

\textbf{Authentication Verification Phase:}  
As detailed in the \textit{message\_authentication\_code\_verification} function of Algorithm~\ref{alg:quantum_authentication}, each verification node $e$ computes the hash value using its local key $k$ and the received digest $D(m)$. If the computed hash matches the one provided in the authentication bundle, the node proceeds to the next consensus step; otherwise, the message is discarded. In cases where the source node $l$ denies the authentication, the verification node $e$ broadcasts a dispute message $\{m, \; m_\alpha, \; 1\}$ and listens for similar reports from other nodes. If $N - f - 1$ consistent dispute messages are received locally, node $l$ is judged to have attempted authentication repudiation.

\subsection{The QDBFT Consensus Scheme}

As illustrated in Fig.~2, the QDBFT scheme operates through five distinct phases: REQUEST, NEW, TRANSMIT, COMMIT, and REPLY, with cryptographic mechanisms applied according to the identities of the communicating participants. For client-to-node communication in the REQUEST and REPLY phases, Post-Quantum Cryptography (PQC) signatures are employed to provide security against quantum-computing-based attacks while maintaining compatibility with standard network interfaces for end-users, whereas for inter-node communication in the NEW, TRANSMIT, and COMMIT phases, the consensus nodes utilize QKD–based Message Authentication (detailed in Section~4.3) to achieve information-theoretically secure identity verification and message integrity during the core consensus process.

\begin{figure}
    \centering
    \includegraphics[scale=0.8]{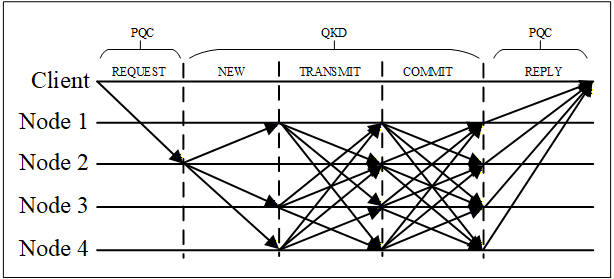}
    \caption{The QDBFT consensus scheme and message flow}
    \label{fig:consensus_process}
\end{figure}

\begin{itemize}
\item[-] \textbf{REQUEST Phase}: The client $\mathcal{C}$ signs the request proposal $m_{\mathrm{REQUEST}}:\{o, t, c\}_\sigma$\footnote{$o$: operation request; $t$: request timestamp; $c$: client identification information.} using a PQC signature scheme and sends it to an arbitrary node (e.g., Node 1 in Fig.~2). Upon receipt, Node 1 stores the proposal in its local buffer pool.

\item[-] \textbf{NEW Phase}: Each node maintains a configuration table $T_v$. 
Through the \textbf{Carousel Mechanism}, the current primary node $\mathcal{P}$ is designated 
(e.g., Node 1 in Fig.~2). Upon entering this phase, a timer is initialized. 
Before the timer expires, $\mathcal{P}$ selects a set of request messages 
$M:\{m_j \mid 0 \leq j \leq G\}$ from the buffer to generate and broadcast 
$m_{\mathrm{NEW}}:\{<i, T_v, fn, H, h, T, D(M), R>, M\}$\footnote{$i$: node ID; $pn$: parent block information; $H$: hash value in the node’s configuration table; $h$: block height; $T$: timestamp for block commitment; $D(M)$: set of message digests $\{D(m_j) \mid 0 \leq j \leq G\}$; $R$: decision result, where 0 denotes rejection and 1 denotes approval.}.  

If $\mathcal{P}$ fails to broadcast $m_{\mathrm{NEW}}$ before the timeout, it is deemed offline. 
In such cases, non-primary nodes $\mathcal{S}$ update the configuration table and reach consensus on $T^u_k$, 
subsequently generating $m_{\mathrm{CurrentPrimary\_missing}}:\{i, j, t, T^u_k\}$\footnote{$j$: the ID of the offline current primary node.}.  
After consensus is achieved, a new primary node $\mathcal{P}^\mathrm{NEW}$ is designated.

\item[-] \textbf{TRANSMIT Phase}: Upon receiving the broadcast from $\mathcal{P}$, non-primary nodes $\mathcal{S}$ verify the authentication tags of $M$ and the decision $R$. They compute the execution result set $R_i:\{r_j \in \{0, 1\} \mid m_j, 1 \leq j \leq \text{Num}(M)\}$ for the messages in $M$. Subsequently, they generate and broadcast $m_\mathrm{TRANSMIT}:\{i, T_v, pn, t, D(m_{\mathrm{NEW}}), D(M), R_i\}$\footnote{$pn$: current primary node information; $R_i$: individual decisions on the primary’s proposals.}.

\item[-] \textbf{COMMIT Phase}: Nodes enter this phase once they have collected $2f+1$ consistent $m_\mathrm{TRANSMIT}$ messages. They then generate and broadcast $m_{\mathrm{COMMIT}}:\{i, T_v, H(pb), T, \text{MerkleTree}\}$\footnote{The Merkle Tree is constructed from the approved proposals.}. Commitment is finalized upon receiving $2f+1$ consistent messages; otherwise, nodes continue processing data from the TRANSMIT phase and complete the commitment once consensus is reached.

\item[-] \textbf{REPLY Phase}: Nodes sign and return the result 
$m_{\mathrm{REPLY}}:\{i, h, t, D(m), r\}_\sigma$\footnote{$r$: final execution result.} 
to the client $\mathcal{C}$. The client accepts the result after receiving $f+1$ consistent responses.
\end{itemize}

\subsection{Scheme Construction}
The QDBFT scheme integrates two complementary cryptographic paradigms to ensure system security. Client-to-node communication is protected by Post-Quantum Cryptography (PQC) signatures, which guarantee integrity, authentication, and resistance to quantum attacks at the network edge. Communication among consensus nodes relies on QKD–based Message Authentication (see Section~4.3), providing information-theoretically secure integrity and authentication. By embedding this quantum authentication directly into the consensus process, the scheme ensures tamper-resistance for all inter-node messages.

\begin{algorithm}
\caption{Carousel: Automatic Node Rotation Algorithm}
\label{alg:carousel}
\SetKwFunction{FSetup}{configuration\_table\_setup}
\SetKwFunction{FSync}{configuration\_table\_sync}
\SetKwFunction{FSelect}{carousel\_node\_selection}
\SetKwFunction{FUpdate}{carousel\_update}

\Fn{\FSetup{$N, d, \{id_i\}$}}{
    \KwIn{$N$: number of nodes; $d$: number of virtual nodes per physical node; $\{id_i\}$: set of node identifiers.}
    \KwOut{$T_v$: initialized configuration table.}
    \For{$i \leftarrow 1$ \KwTo $N$}{
        \For{$q \leftarrow 1$ \KwTo $d$}{
            $T_v[id_i] \leftarrow T_v[id_i] \cup \{ (2^{32}/i) \times q \bmod 2^{32} \text{ as a hash point} \}$\;
        }
    }
    \textbf{Broadcast} $m: \{T_v, m_\alpha\}$\;
    \If{Received $2f$ consistent $T_v$ and valid $m_\alpha$}{
        \Return $T_v$\;
    }
}

\Fn{\FSelect{$T_v, H(pb)$}}{
    \KwIn{$T_v$: current configuration table; $H(pb)$: hash of the parent block.}
    \KwOut{$id$: identifier of the selected primary node.}
    $target \leftarrow H_{32}(H(pb))$\;
    Find the smallest hash point $H_z$ in $T_v$ such that $H_z > target$\;
    \Return the node $id$ corresponding to $H_z$\;
}

\Fn{\FUpdate{$T_v, m$}}{
    \KwIn{$T_v$: current configuration table; $m$: control message (Join $m_{jo}$, Exit $m_{exit}$, or Unresponsive $m_{unresp}$).}
    \Case{$m$ is $m_{jo}: \{j, t, T'_v\}$ (Node Join)}{
        \If{$T'_v \geq T_v$}{
            \textbf{Broadcast} $m_{agr}: \{i, j, t, T'_v, D(m_{jo})\}$\;
            \If{Received $2f$ consistent $m_{agr}$}{
                \textbf{Broadcast} $m_{agr\_c}$\;
                \If{Received $2f$ consistent $m_{agr\_c}$}{
                    Update $T_v \rightarrow T^u_v$ and \Return $T^u_v$\;
                }
            }
        }
    }

    \Case{$m$ is $m_{exit}: \{j, t, T'_v\}$ (Node Exit)}{
    \textbf{Broadcast} $m_{exit\_broad}: \{i, j, t, T'_v, D(m_{exit})\}$\;
    \If{Received $(2f-1)$ consistent $\{m_{exit}\ \text{or}\ m_{exit\_broad}\}$}{
        \textbf{Broadcast} $m_{lc}$\;
        \If{Received $(2f-1)$ consistent $m_{lc}$}{
            Update $T_v \rightarrow T^u_v$\;
        }
    }
}

    \Case{$m$ is $m_{unresp}: \{l, j, t, T'_v\}$ (Node Unresponsive)}{
        Send $m_{sur}$ (survival probe) to node $j$\;
        \If{No response within threshold $\Delta T$}{
            \textbf{Broadcast} $m_{unresp}: \{i, j, t, T^u_v\}$\;
            \If{Received $2f-1$ consistent $m_{unresp}$}{
                Update $T_v \rightarrow T^u_v$\;
            }
        }
    }
}
\end{algorithm}

The Carousel mechanism is detailed in Algorithm~2. In the \textit{configuration\_table\_setup} function, nodes apply the $H_{32}(\cdot)$ function to compute the distribution of virtual nodes and generate the initial configuration table $T_v$, which is stored locally once consensus is reached (i.e., upon receiving $2f$ consistent messages, forming the $2f+1$ quorum). The \textit{configuration\_table\_sync} function updates the local table when a node receives $2f$ consistent update messages from the consensus network. In the \textit{carousel\_node} function, each node independently determines the next primary identifier using the parent block hash. Finally, the \textit{carousel\_update} function manages dynamic membership changes: node joining is confirmed after $2f$ consistent $m_{jo}$ messages; node leaving is processed after $2f-1$ consistent $m_{exit}$ messages; and node unresponsive detection is triggered by $m_{unresp}$ messages, verified through survival probes before broadcasting and updating $T_v$ accordingly.

\begin{algorithm}
\caption{Client-Node Interaction Algorithm}
\label{alg:interaction}

\SetKwFunction{FReq}{request}
\SetKwFunction{FRep}{reply}

\Fn{\FReq{}}{
    \textbf{Client $\mathcal{C}$:}\;
    \Indp
    Generate $m_{\mathrm{REQUEST}} \leftarrow \{o, t, c\}_\sigma$ using PQC signature\;
    Randomly send $m_{\mathrm{REQUEST}}$ to an arbitrary node $l \in N$\;
    \Indm
    \textbf{All Nodes:}\;
    \Indp
    Verify PQC signature $\sigma$\;
    \If{Verification successful}{
        Add $m_{\mathrm{REQUEST}}$ to the local transaction \textit{pool}\;
    }
    \Indm
    \textbf{Primary Node $\mathcal{P}$:}\;
    \Indp
    Identify the current $\mathcal{P}$ via the \textit{carousel\_node} function\;
    Select $G$ transactions from the \textit{pool} to initiate the \textbf{NEW} phase\;
    \Indm
}

\vspace{0.5em}

\Fn{\FRep{}}{
    \textbf{All Nodes:}\;
    \Indp
    Generate $m_{\mathrm{REPLY}}: \{i, h, t, D(m), r\}_\sigma$\
    Send $m_{\mathrm{REPLY}}$ back to the client $\mathcal{C}$\;
    \Indm
    \textbf{Client $\mathcal{C}$:}\;
    \Indp
    \If{Received $f+1$ consistent $m_{\mathrm{REPLY}}$ messages}{
        Verify each PQC signature $\sigma$\;
        \If{Signatures are valid}{
            Accept the result and finalize the request\;
        }
    }
    \Indm
}
\end{algorithm}

The interaction between nodes and the client is detailed in Algorithm~3. In the \textit{request} function, the client employs a Post-Quantum Cryptography (PQC) digital signature scheme to sign the operation $o$ intended for inclusion in the blockchain. The signed request $\sigma$ is then randomly dispatched to a node $l \in N$. Upon receipt, node $l$ verifies the validity of $\sigma$; if authenticated, the request is inserted into the local proposal buffer pool. When node $l$ is designated as the current primary via the \textit{carousel\_node} function, it selects $G$ operation requests from the buffer pool and initiates the \textbf{NEW} phase. In the \textit{reply} function, following the completion of the \textbf{COMMIT} phase, all nodes generate and return the $m_{\mathrm{REPLY}}$ message to the client. Once the client successfully verifies the PQC signatures, the operation is formally recognized as committed to the blockchain.

\begin{algorithm}
\caption{Inter-Node Consensus Interaction Algorithm}
\label{alg:node_interaction_full}
\SetKwFunction{FNew}{new}
\SetKwFunction{FTrans}{transmit}
\SetKwFunction{FComm}{commit}
\SetKwFunction{FReply}{reply}
\SetKwFunction{FSelect}{carousel\_node}
\SetKwFunction{FUpdate}{carousel\_update}

\Fn{\FNew{}}{
    \textbf{Primary Node $\mathcal{P}$:}\;
    \Indp
    $\mathcal{P} \leftarrow \FSelect{}$\;
    Select $G$ requests $M:\{m_j\}$ from the pool and compute digests $D(M)$\;
    Process $M$ to determine decision results $R:\{r_j \in \{0, 1\}\}$\;
    \textbf{Broadcast} $m_{\mathrm{NEW}}: \{ \langle i, T_v, fn, H(pb), h, T, D(M), R \rangle, M \}$\;
    \Indm
    \textbf{Other Nodes (Followers) $\mathcal{S}$:}\;
    \Indp
    \If{Received $m_{\mathrm{NEW}}$}{
        Execute \FTrans{}\;
    }
    \ElseIf{Timer $T$ exceeds threshold $\Delta T$}{
        $T^u_k \leftarrow \FUpdate{}$\;
        Elect $\mathcal{P}^{\mathrm{NEW}}$ based on $H_{32}(H(pb))$ and restart \FNew{}\;
    }
    \Indm
}

\vspace{0.5em}

\Fn{\FTrans{}}{
    \textbf{Follower Node $i$:}\;
    \Indp
    \If{Received $m_{\mathrm{NEW}}$}{
        $R_i \leftarrow$ Verify $M$ and $R$ to generate local decision results\;
        \textbf{Broadcast} $m_{\mathrm{TRANSMIT}}: \{i, T_v, primary\_id, t, D(m_{\mathrm{NEW}}), D(M), R_i\}$\;
    }
    \If{Received $2f$ consistent $m_{\mathrm{TRANSMIT}}$}{
        Execute \FComm{}\;
    }
    \Indm
}

\vspace{0.5em}

\Fn{\FComm{}}{
    \textbf{All Nodes:}\;
    \Indp
    Construct a Merkle Tree from approved transactions\;
    \textbf{Broadcast} $m_{\mathrm{COMMIT}}: \{i, T_v, H(pb), T, \text{MerkleTree}\}$\;
    \If{Received $2f$ consistent $m_{\mathrm{COMMIT}}$}{
        Execute \FReply{}\;
    }
    \Indm
}

\vspace{0.5em}

\Fn{\FReply{}}{
    \textbf{All Nodes:}\;
    \Indp
    Generate $m_{\mathrm{REPLY}}: \{i, h, t, D(M), R\}_\alpha$\;
    Send $m_{\mathrm{REPLY}}$ back to the client $\mathcal{C}$\;
    \Indm
}
\end{algorithm}

The interaction between nodes is detailed in Algorithm~4. 
\begin{itemize}
    \item \textbf{NEW Phase}: The current primary node is designated via the \textit{carousel\_node} mechanism. It selects $G$ proposal requests ($m_{\mathrm{REQUEST}}$) from its local buffer pool, generates digests for each proposal, and determines the initial decision results $r$. The node then broadcasts the $m_{\mathrm{NEW}}$ message and transitions to the \textit{transmit} phase. If a timeout $\Delta T$ occurs and follower nodes have not received $m_{\mathrm{NEW}}$, the \textit{carousel\_update} algorithm is invoked to reach consensus on updating the configuration table to $T^u_k$. After electing a new primary node, the system re-enters the \textbf{NEW} phase.

    \item \textbf{TRANSMIT Phase}: When a non-primary node $i$ receives $m_{\mathrm{NEW}}$ and successfully verifies it, the node processes the decision information $R$ to generate its own local decision result $R_i$. Node $i$ then broadcasts the $m_{\mathrm{TRANSMIT}}$ message. Once a node receives $2f$ consistent $m_{\mathrm{TRANSMIT}}$ messages (forming a $2f+1$ quorum including its own), it proceeds to the \textbf{COMMIT} phase.

    \item \textbf{COMMIT Phase}: All nodes broadcast the $m_{\mathrm{COMMIT}}$ message and verify incoming commitment messages from others. If a node collects $2f$ consistent $m_{\mathrm{COMMIT}}$ messages, the block is formally committed to the ledger, and the system transitions to the \textbf{REPLY} phase.
\end{itemize}

\section{Analysis} 
\subsection{Security Analysis} 
This section evaluates the correctness and security of the proposed consensus algorithm, defines the relevant properties, and provides a formal security analysis of the scheme.

\noindent \textbf{Theorem 1.} \textit{System Liveness.} The system can distinguish between malicious and honest client requests and formulate independent judgments. Consensus is finalized only when $2f+1$ consistent judgments are reached. Furthermore, the configuration table is dynamically updated based on the performance of primary nodes to maintain system health and mitigate potential damage from malicious participants.

\noindent \textit{Proof.} For an honest client $\mathcal{C}$, a legitimate proposal $m_{\mathrm{REQUEST}}:\{o, t, c\}_\sigma$ is stored in the buffer pool after passing initial verification. Within the consensus timeout, this proposal is validated by at least $2f+1$ honest nodes, reaching consensus and final commitment, with the response $m_{\mathrm{REPLY}}:\{i, h, t, D(m), r\}_\sigma$ returned to $\mathcal{C}$. Conversely, malicious requests from dishonest clients fail the validity checks of honest nodes and are rejected. If a dishonest client colludes with a malicious primary node to propose an invalid request, the honest majority will generate $2f+1$ consistent rejection decisions during the \textbf{TRANSMIT} phase, thereby preventing the request from being committed to the ledger.

Regarding network dynamics, node joins, departures, and unresponsive events trigger a consensus-based update of the configuration table $T_v$. During the \textbf{NEW} phase, if a primary node becomes unresponsive, the system incurs only a delay of $\Delta T$ before proposal rights are transferred to the next healthy node according to $T_v$. For proposals that have already passed the \textbf{TRANSMIT} phase, the \textbf{COMMIT} phase can still be successfully finalized via the $T_v$ mechanism even if the current primary node fails.

The Carousel mechanism applies the $H_{32}(\cdot)$ algorithm to map the parent block’s hash onto the hash ring, ensuring that each node has an equal probability of obtaining the rights to propose the next block. Malicious behaviors by a primary node (e.g., throttling throughput or increasing latency) can only affect the current block. This prevents malicious nodes from retaining leadership for extended periods, thereby avoiding long-term inefficiency or systemic failure.

\noindent \textbf{Theorem 2.} 
\textit{Block Consistency.} All honest nodes maintain strict synchronization of the global blockchain state. The data stored by each node prior to a checkpoint is fully consistent, and any information broadcast by an honest node is eventually recorded and synchronized by all other honest nodes.

\noindent \textit{Proof.} When $m_{\mathrm{CHECKPOINT}}$ messages are broadcast and consensus is reached, nodes verify the validity of all message logs preceding the checkpoint. This guarantees that messages prior to the checkpoint remain strictly consistent and that the parent block hash in the block header is correct. For any client transaction proposal issued by a legitimate primary node, $2f+1$ consistent $m_{\mathrm{COMMIT}}$ messages are required for successful inclusion in a block. The corresponding log information is then securely replicated across the network. Because the $2f+1$ quorum necessarily includes at least $f+1$ honest nodes, no two conflicting blocks can be committed at the same height, thereby preserving the integrity of the global state.

\noindent \textbf{Theorem 3.} \textit{Consistent Delivery \cite{9833787}.} The system guarantees that a request submitted by an honest client will eventually receive a response that is strictly consistent with the system's execution under a specific configuration state.

\noindent \textit{Proof.} Consistent delivery is ensured through the globally ordered processing and state transition mechanisms of State Machine Replication (SMR) \cite{SMR1978, SMR1984, SMR1990}. This framework guarantees both the liveness and consistency of requests during membership changes. Specifically, the liveness analysis is provided in Theorem~1, and the consistency analysis is detailed in Theorem~2. By relying on the real-time updates of the configuration table $T_v$ and the $2f+1$ consistent decision quorum, the system ensures that proposals from honest clients will eventually reach consensus, be packaged into blocks, and receive an honest response. Furthermore, if a node that completed message submission during the \textbf{TRANSMIT} phase becomes unresponsive during the \textbf{COMMIT} phase, it can recover and synchronize the block log information from healthy nodes upon rejoining the network.

\noindent \textbf{Theorem 4.} \textit{Enhanced Total Ordering \cite{9833787}.} 
If nodes with different versions of the configuration table deliver a request with the same sequence number, the content of these requests must be identical.

\noindent \textit{Proof.} Building upon the total ordering property, the system requires all honest nodes to reach strict agreement on the global transaction sequence. This guarantees linearizable execution and state synchronization even under network asynchrony, dynamic membership changes, or malicious interference. The sequence number of a client’s proposal request is uniquely determined by the nodes. When a client broadcasts a proposal $m_{\mathrm{REQUEST}}$, all nodes verify its validity before storing it in their local buffer pools. When the current primary node selects this request to initiate the \textbf{NEW} phase, follower nodes first verify that their local configuration table $T_v$ is up to date. They then check for the existence of the corresponding request in their buffer pools and remove it once processed. Any message associated with an outdated version of $T_v$ within the same consensus stage is rejected. This mechanism ensures that nodes, regardless of their configuration state, only commit identical requests for any given sequence number.

\noindent \textbf{Theorem 5.} \textit{Quantum Resistance.} The cryptographic primitives integrated into the consensus algorithm are resilient against threats posed by quantum computing.

\noindent \textit{Proof.} The quantum resistance of the QDBFT scheme relies on Post-Quantum Cryptography (PQC) and one-time hash functions. 
\begin{itemize}
    \item \textbf{Client-to-Node Security:} The system employs PQC algorithms for digital signatures between clients and nodes. This design is modular, allowing for the flexible integration of more efficient PQC signature schemes as they emerge in the future.
    \item \textbf{Inter-Node Authentication:} Consensus messages exchanged between nodes are authenticated using one-time hash functions combined with QKD-based key distribution, ensuring resilience against quantum adversaries.
    \item \textbf{Forward Security:} Even if a node’s long-term key material is compromised, previously committed blocks remain secure due to the one-time nature of hash-based authentication and the immutability of the blockchain ledger.
\end{itemize}

\begin{itemize} \item \textbf{Inter-Node Security:} Inter-node communication is protected by \textbf{QKD-based authentication}. This mechanism leverages information-theoretically secure (ITS) symmetric keys established via a Quantum Key Distribution (QKD) network. The security level can be dynamically configured according to performance requirements: by employing a one-time hash algorithm and refreshing keys after each authentication, the scheme achieves \textbf{ITS-grade authentication} (consistent with the One-Time Pad principle). Alternatively, to enhance practical efficiency, the scheme supports \textbf{computationally secure authentication} by applying HMAC to message digests with periodic key updates. This dual-mode design ensures robustness against quantum adversaries—such as those exploiting Grover's algorithm—while maintaining high consensus throughput. \end{itemize} \subsection{Efficiency Analysis} 
The efficiency of the proposed system was evaluated on a computing platform equipped with an Apple M1 chip, featuring high-performance cores operating at a clock speed of 3.2 GHz. The system is further supported by 16 GB of unified memory, providing sufficient computational and memory resources to sustain high-throughput consensus operations under realistic workloads.

\subsubsection{Authentication Overhead}

The authentication for messages exchanged between nodes utilizes QKD-based authentication. To maintain the highest security standards, QKD-derived keys are immediately discarded following a single authentication operation. Consequently, for a network size of $N$ nodes, the key consumption across the \textbf{NEW}, \textbf{TRANSMIT}, and \textbf{COMMIT} phases is summarized in Table~2. Under normal operating conditions, the total key consumption for a single complete consensus round follows the rule $2N(N-1)$ units.

For QKD-based authentication, the Toeplitz hash algorithm \cite{bernstein2005poly1305} is implemented in our system and benchmarked against HMAC \cite{bellare1996keying}, ECDSA \cite{johnson2001elliptic}, RSA \cite{rivest1978method}, and the post-quantum algorithm Falcon-512 \cite{fouque2018falcon}. As illustrated in Fig.3a, when the number of participating nodes $N$ increases from 4 to 10, only the authentication generation times for Toeplitz and HMAC exhibit a linear upward trend. This is attributed to the fact that the size of the authentication values generated by Toeplitz and HMAC is proportional to the number of participating nodes, whereas the authentication sizes of the other algorithms remain independent of $N$. Despite this, the authentication rate of the Toeplitz algorithm remains significantly higher than that of RSA.

Regarding authentication verification speed, as shown in Fig.~3b, HMAC achieves the highest throughput. The verification rate of the Toeplitz algorithm is notably higher than that of ECDSA, demonstrating its efficiency for high-frequency consensus operations.

\begin{table}[b]
\caption{Key Consumption per Consensus Round}
\label{tab:key_consumption_data}
\centering
\begin{tabular}{cc} 
\toprule
\textbf{Number of Nodes ($N$)} & \textbf{Key Consumption (Units)} \\ 
\midrule
4  & 24   \\
5  & 40   \\
6  & 60   \\
7  & 84   \\
8  & 112  \\
9  & 144  \\
10 & 180  \\
\bottomrule
\end{tabular}
\end{table}

\begin{figure*}
    \centering
    \begin{minipage}{0.48\textwidth}
        \centering
        \includegraphics[scale=0.25]{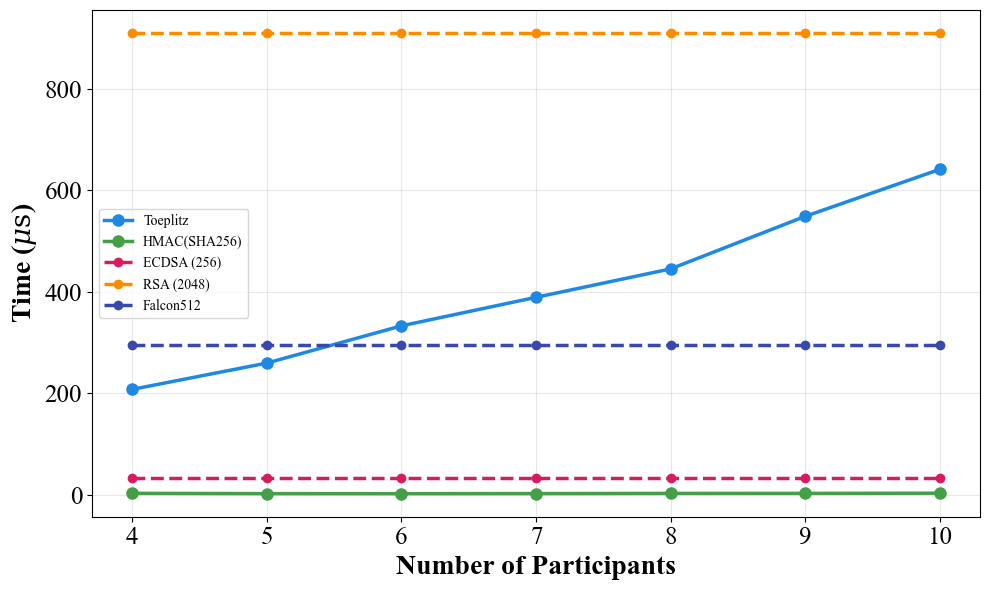}
        \\
        \vspace{0.2cm}
        \small{(a). Authentication and signature generation time across different node counts}
    \end{minipage}
    \hfill
    \begin{minipage}{0.48\textwidth}
        \centering
        \includegraphics[scale=0.25]{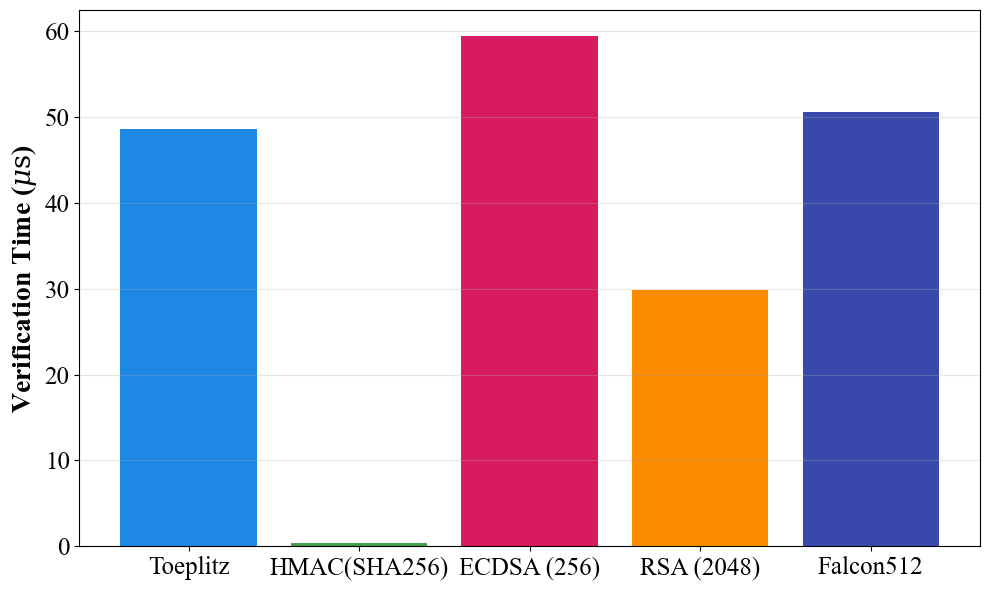}
        \\
        \vspace{0.2cm}
        \small{(b). Authentication and signature verification time across different node counts}
    \end{minipage}
    \\
    \vspace{0.3cm}
    \caption{Performance evaluation of signature and verification processes}
    \label{fig:performance_analysis}
\end{figure*}

In summary, the choice of authentication algorithm provides a flexible balance between throughput and security levels. If the primary objective is to maximize the on-chain data rate, the HMAC algorithm serves as the optimal option due to its superior verification efficiency. Conversely, for future-oriented applications where the long-term integrity of authentication data is paramount, the Toeplitz algorithm is required to provide robust protection against evolving quantum threats, thereby ensuring information-theoretic security throughout the consensus process.

\subsubsection{Consensus Mechanism Efficiency}
The efficiency of the consensus mechanism was evaluated within a laboratory network environment, where real-world network conditions were simulated by controlling the network latency. We benchmarked the Transactions Per Second (TPS) of the QDBFT and PBFT algorithms across configurations of 4, 7, and 10 nodes.As illustrated in Fig. 4a and Fig. 4b, the experimental results for simple transaction transfers show that the TPS performance of QDBFT is nearly identical to that of PBFT. This similarity arises because the time consumption of the Toeplitz algorithm (used in QDBFT) and the ECDSA algorithm (used in PBFT) differs only at the microsecond ($\mu s$) level. Compared to the millisecond ($ms$) level of network latency, these computational overheads are negligible. Consequently, under normal operating conditions without node failures, the TPS of both algorithms decreases as network latency increases. Furthermore, under the same network latency, an increase in the number of participants leads to a reduction in TPS.

Additionally, the duration of a single consensus round was measured (calculated based on the transaction volume per minute), as shown in Fig. 4c and Fig. 4d. The results indicate that the consensus latency of QDBFT and PBFT remains highly consistent, with the total consensus time being directly proportional to the network latency. Specifically, the time required for one consensus round is approximately three times the network latency. This corresponds to the cumulative delay incurred during the three essential phases—\textbf{NEW}, \textbf{TRANSMIT}, and \textbf{COMMIT}—plus the marginal time for authentication generation and verification. Consistent with the TPS findings, the time consumed for a single consensus round increases with the number of participating nodes.

\begin{figure*}
    \centering
    \begin{minipage}{0.48\textwidth}
        \centering
        \includegraphics[width=\linewidth]{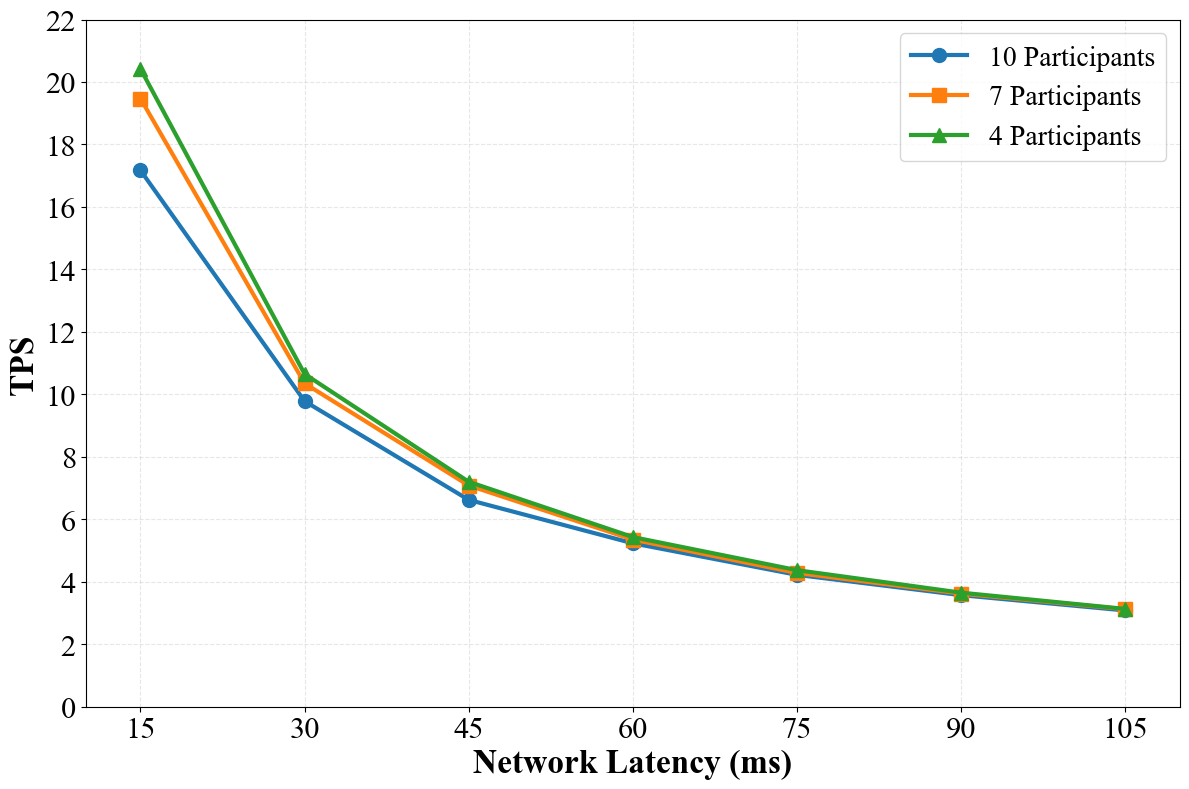}
        \\ \vspace{0.1cm}
        \scriptsize{(a). TPS performance of QDBFT}
    \end{minipage}
    \hfill
    \begin{minipage}{0.48\textwidth}
        \centering
        \includegraphics[width=\linewidth]{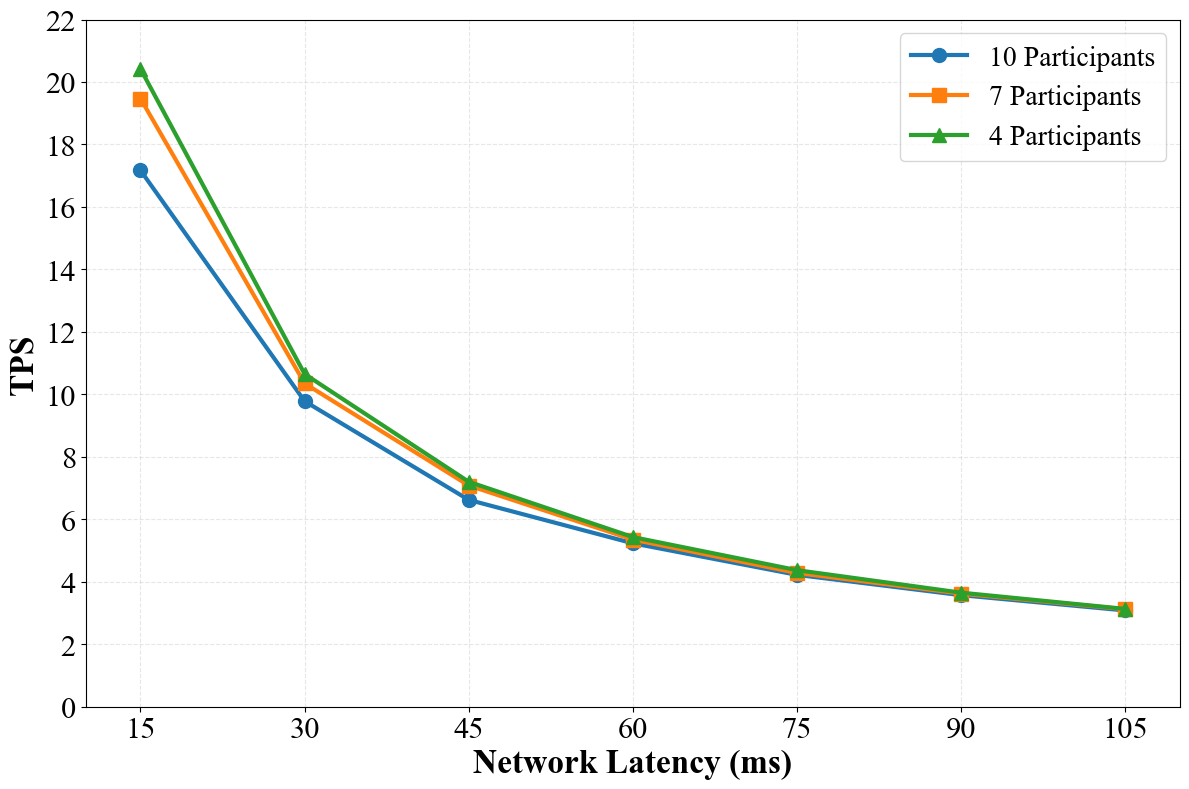}
        \\ \vspace{0.1cm}
        \scriptsize{(b). TPS performance of PBFT \\ \ }
    \end{minipage}
    \vspace{0.5cm}
    
    \begin{minipage}{0.48\textwidth}
        \centering
        \includegraphics[width=\linewidth]{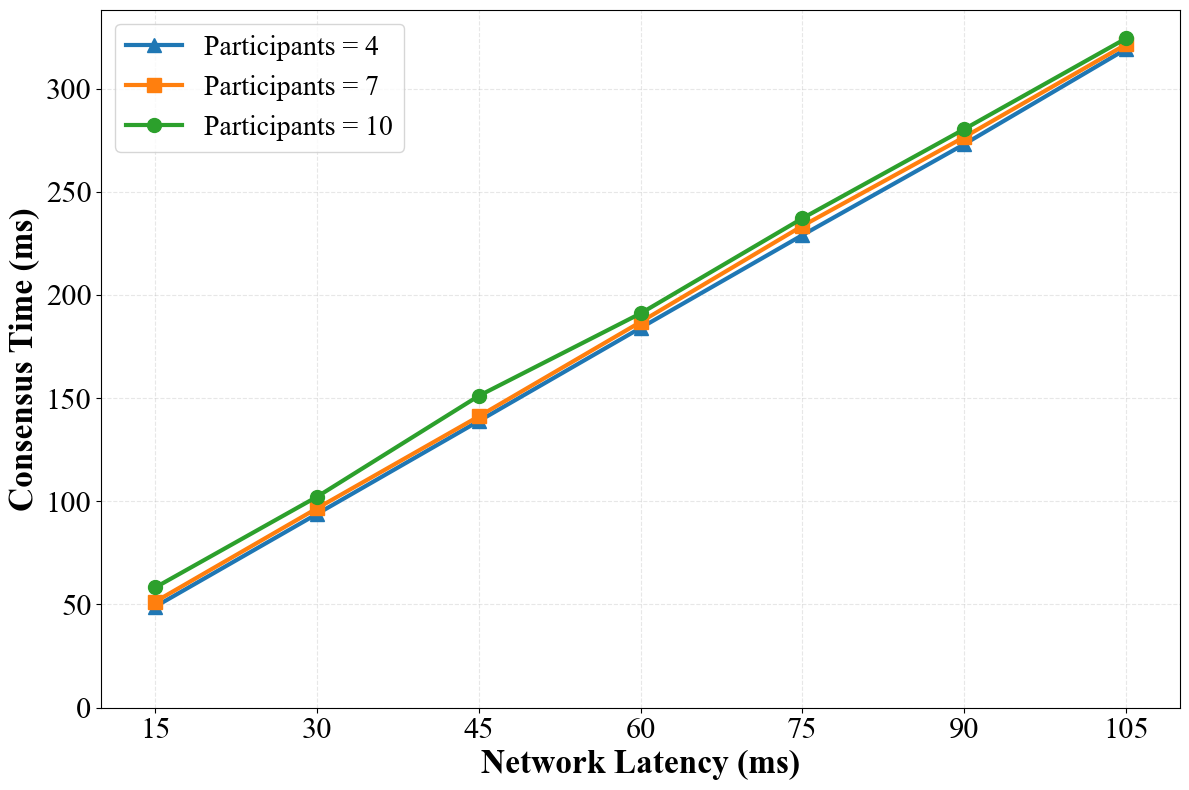}
        \\ \vspace{0.1cm}
        \scriptsize{(c). Consensus latency of QDBFT}
    \end{minipage}
    \hfill
    \begin{minipage}{0.48\textwidth}
        \centering
        \includegraphics[width=\linewidth]{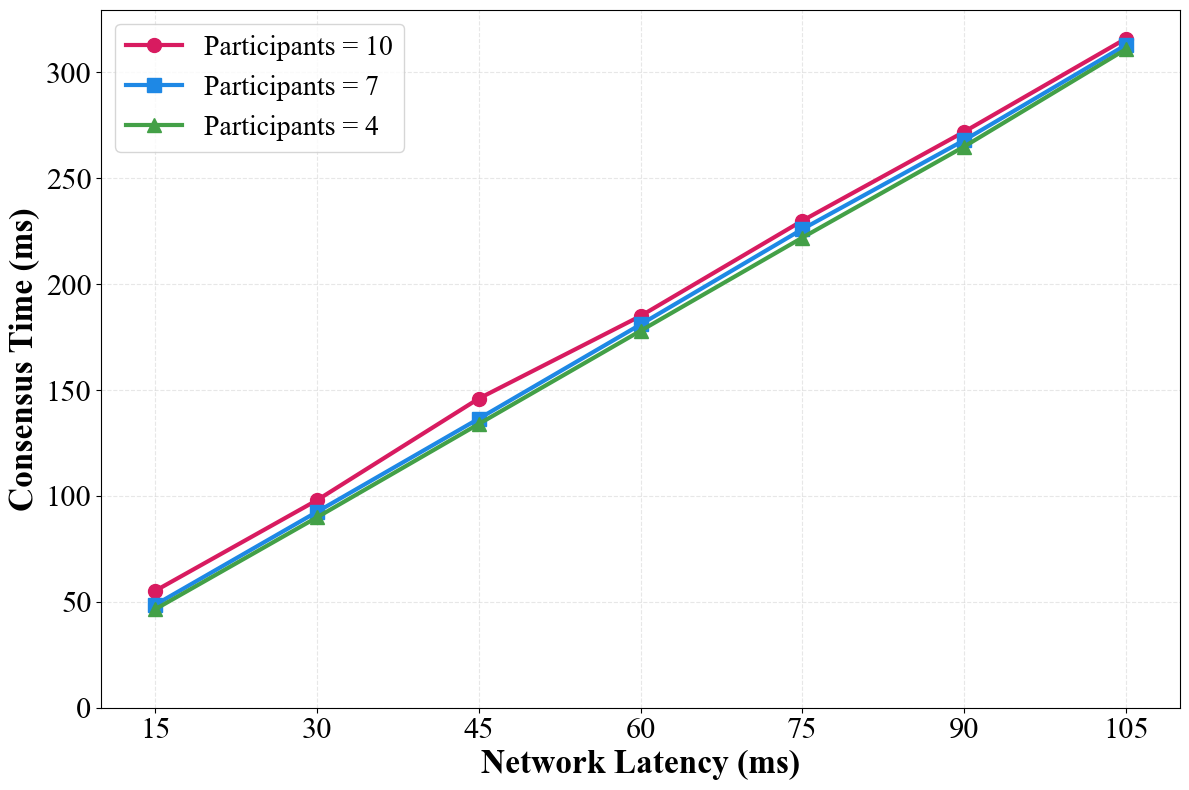}
        \\ \vspace{0.1cm}
        \scriptsize{(d). Consensus latency of PBFT}
    \end{minipage}
    
    \vspace{0.3cm}
    \caption{Efficiency comparison between QDBFT and PBFT in terms of TPS and consensus latency across varying node scales and network delays.}
    \label{fig:efficiency_comparison}
\end{figure*}

\subsubsection{Dynamic Node Reconfiguration}
Based on the automatic primary node rotation mechanism proposed in this work, any operation involving the addition, withdrawal, or offline status of a node triggers a reconfiguration consensus process. 

Once the reconfiguration process begins, the blockchain commitment is temporarily suspended and resumes only after the reconfiguration is completed. As illustrated in Fig. 5a and Fig. 5b, under identical network latency, the time required for node addition or withdrawal increases with the number of participating nodes. As shown in Fig. 5c, when a primary node becomes unresponsive, the remaining nodes monitor its status via heartbeat packets. If the node remains inactive after a 5-second timeout, a withdrawal procedure is initiated. Once consensus is reached on the addition, withdrawal, or offline status of a node, a new configuration table $T_v$ is generated. Subsequently, all operational nodes elect the new primary node based on the updated configuration in $T_v$, thereby ensuring continuous and stable system operation.

\begin{figure*}[htbp]
    \centering
    \begin{minipage}{0.31\textwidth}
        \centering
        \includegraphics[width=\linewidth]{figures/E-Consensus-QDBFT.png}
        \\ \vspace{0.1cm}
        \scriptsize{(a). Latency when adding a new node \\ \ }
    \end{minipage}
    \hfill
    \begin{minipage}{0.31\textwidth}
        \centering
        \includegraphics[width=\linewidth]{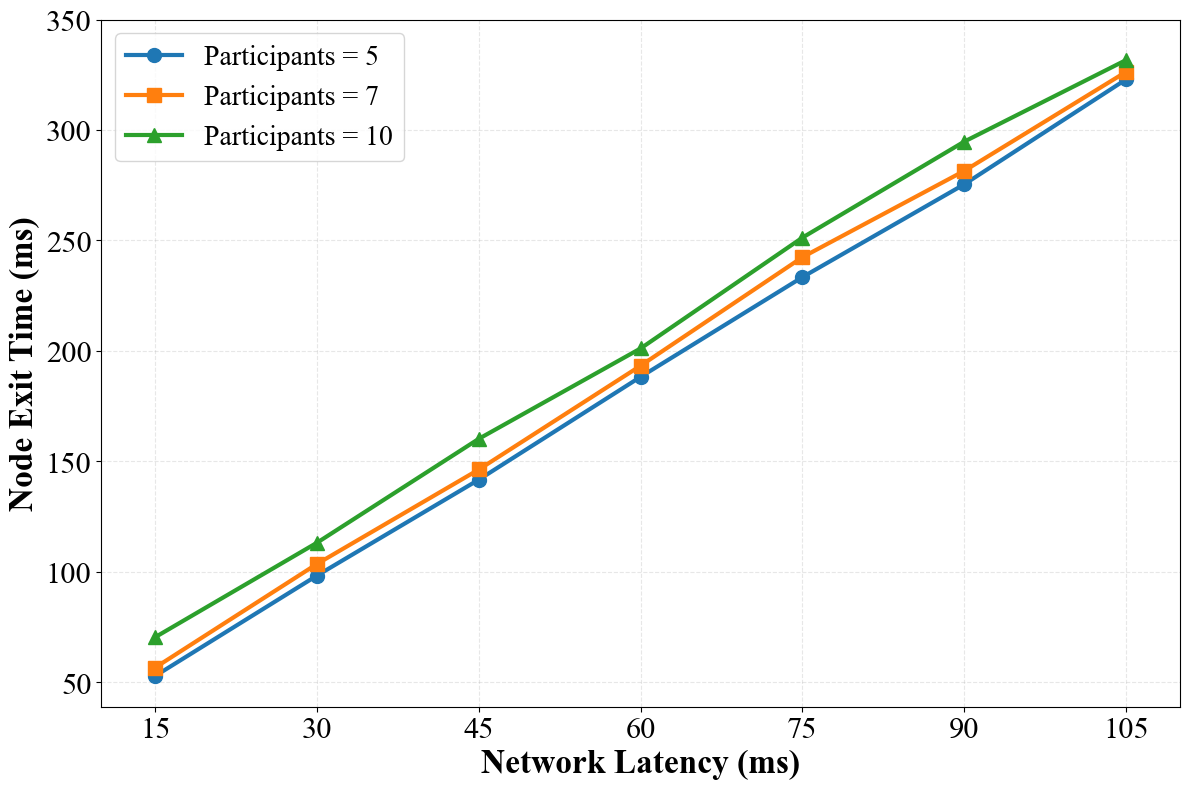}
        \\ \vspace{0.1cm}
        \scriptsize{(b).  Latency when a node exits \\ \ }
    \end{minipage}
    \hfill
    \begin{minipage}{0.31\textwidth}
        \centering
        \includegraphics[width=\linewidth]{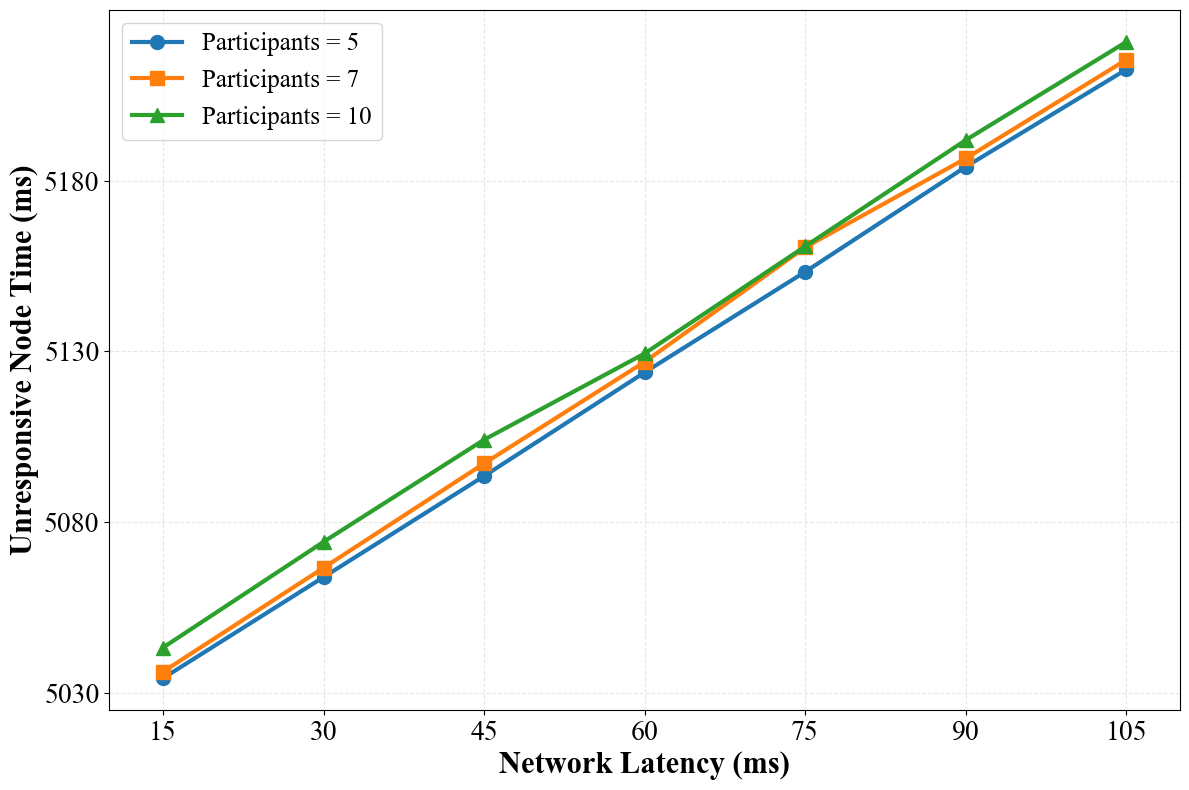}
        \\ \vspace{0.1cm}
        \scriptsize{(c). Latency when a node becomes unresponsive}
    \end{minipage}
    
    \vspace{0.3cm}
    \caption{Time overhead analysis for dynamic node reconfiguration scenarios.}
    \label{fig:node_reconfiguration}
\end{figure*}

\subsection{Discussion}
In blockchain systems, nodes store all historical block data locally. As the chain length grows, the demand for local storage increases rapidly. To improve storage efficiency, the QDBFT scheme incorporates a checkpoint mechanism similar to PBFT \cite{castro1999practical,castro1999authenticated}. At initialization, the system defines a block threshold; once this threshold is reached, any node may broadcast an $m_{\mathrm{CHECKPOINT}}$ message. After consensus on the checkpoint, nodes prune non-essential historical data prior to the checkpoint, retaining only block headers and state information.

QDBFT extends traditional BFT architectures with several innovations. Most notably, conventional public-key signatures are replaced with \textbf{QKD-based authentication} for inter-node communication. For client–node interactions, post-quantum cryptography (PQC) standards such as ML-DSA \cite{NIST:FIPS204} and SLH-DSA \cite{nist_fips_205} are employed. The QKD key consumption per consensus round is quantified in Table~2. Compared to ECDSA in classical BFT protocols, the QKD-based authentication employed here is particularly well-suited for consortium blockchain systems with committee sizes of up to 10 nodes. Thus, QDBFT offers a quantum-resistant alternative for message authentication in decentralized systems. For larger committees or higher participation, the HMAC algorithm can serve as a high-efficiency substitute for Toeplitz, providing a practical balance between performance and security requirements.

\bmhead{Acknowledgements}

This research was primarily funded by the "Soft Science Research" project under the 2025 High-Level Institutional Development and Operation Program of Shanghai, People's Republic of China (No. 25692107800), the Guangdong Provincial Key Area R\&D Program (No. 2020B03\\03010001), the Quantum Science and Technology-National Science and Technology Major Project (No. 2021ZD0301300), the National Wide-Area Quantum Secure Communication Backbone Network Project, and the "Eastern Data, Western Computing" Demonstration Project: Quantum Trusted Cloud Project.  We express our sincere gratitude for the support and assistance from these project funds.


\bibliography{sn-bibliography}

\end{document}